# HIGHWAY MOBILITY AND VEHICULAR AD-HOC NETWORKS IN NS-3

Hadi Arbabi
Michele C. Weigle

Department of Computer Science
Old Dominion University
Norfolk, VA 23529, USA

## ABSTRACT

The study of vehicular ad-hoc networks (VANETs) requires efficient and accurate simulation tools. As the mobility of vehicles and driver behavior can be affected by network messages, these tools must include a vehicle mobility model integrated with a quality network simulator. We present the first implementation of a well-known vehicle mobility model to ns-3, the next generation of the popular ns-2 networking simulator. Vehicle mobility and network communication are integrated through events. User-created event handlers can send network messages or alter vehicle mobility each time a network message is received and each time vehicle mobility is updated by the model. To aid in creating simulations, we have implemented a straight highway model that manages vehicle mobility, while allowing for various user customizations. We show that the results of our implementation of the mobility model matches that of the model's author and provide an example of using our implementation in ns-3.

## 1 INTRODUCTION

Vehicular ad-hoc networks (VANETs) are networks in which each node is a vehicle. Such systems aim to provide communications between individual vehicles and between vehicles and nearby fixed equipment, or roadside units. The goal of VANETs, and more broadly vehicular networks, is to improve traffic safety by providing timely information to drivers and concerned authorities. The development of VANETs has received much attention from the automotive industry and government agencies, including the US Department of Transportation (DOT) which has launched the IntelliDrive initiative (US-DOT 2010). The US DOT reports that in 2008, 37,000 people died in traffic accidents in the US. The agency sees the promise of IntelliDrive, and VANETs in general, to be able to significantly reduce that number.

In order to provide applications that can fulfill this vision, approaches must be thoroughly evaluated. There are a limited number of testbeds with instrumented vehicles and roadside units. As this is prohibitively expensive for most academic researchers, the majority of evaluation studies have been performed via simulation. VANET simulations have typically been segregated into traffic simulations and network simulations. Traffic simulators, such as CORSIM (Halati et al. 1997), SUMO (Krajzewicz et al. 2006), VISSIM (PTV America 2010), and VanetMobiSim (Fiore et al. 2006) have been used to generate realistic mobility traces of vehicle traffic. These traces would then be fed into well-known network simulators such as ns-2 (Breslau et al. 2000), QualNet (Scalable Network Technologies 2010), OPNET (2010), or GloMoSim (Zeng et al. 1998) to measure network performance. VANET tools such as TraNS (Piorkowski et al. 2008) and MOVE (Karnadi et al. 2007) have been used to facilitate this interaction between traffic and network simulators. More recently, researchers have developed integrated simulators such as ASH (Ibrahim and Weigle 2008) and Gorgorin et al. (2006) that allow feedback between the applications using the network and the traffic model. This is important because the goal of most VANET applications is to provide drivers with information that may change their driving behavior or allow them to make more in-formed decisions (*e.g.*, start braking now, or take the next exit to avoid a traffic jam). Interested readers can find detailed comparisons of various VANET simulators in Hassan (2009) and Yan et al. (2009).

The problem with integrated simulators is that often either the mobility model is overly simplified or the network model is overly simplified. In order to study important networking properties of VANETs, a high quality network simulator is essential. We have chosen to balance these two concerns by taking the latest version of the highly-regarded network simulator, ns-3 (Henderson et al. 2006), and adding a well-known traffic mobility model in order to provide an integrated simulator for VANET research. ns-3 is a discrete-event network simulator written in C++, targeted primarily for research and educational use, and intended as a replacement for the popular ns-2 simulator. ns-3 promises to be a more efficient and more accurate simulator than its predecessor (especially for wireless protocols). In addition, during the first quarter of 2010, ns-3 averaged almost 7000 downloads per month (<http://www.nsnam.org>). For this reason, we were interested in using ns-3 to perform our VANET simulations. ns-3 provides various mobility models, but none are appropriate to simulate the mobility of vehicles. The mobility of a node in the mobility models included in ns-3 depends only on the node itself. In realistic vehicular mobility, the mobility of the node must depend on the surrounding nodes and the conditions on the road. Furthermore, this node dependency becomes essential when messages in the network can affect the mobility of the nodes on the roads. For example, the receipt of a safety message may result in a speed reduction. Fiore and Harri (2008) and Fiore (2009) investigated the effects of node mobility on network characteristics. They found that realistic mobility, especially at intersections, has a great impact on networking connectivity metrics and that car-following models, such as the Intelligent Driver Model (IDM) (Treiber et al. 2000), provide realistic movement. In addition, they found that multi-lane scenarios are important when considering network-level clustering.

We have implemented IDM and the MOBIL lane change model (Treiber and Helbing 2002) in ns-3. In addition, we have provided a *Highway* class to represent a straight multi-lane, bi-directional roadway. In our simulations, the *Highway* object is the "brain" of the system and efficiently manages the behavior of vehicles and their mobility on the road. Each vehicle is a fully-fledged wireless node in ns-3. In this way, vehicles can move with realistic mobility and communicate with each other to form a VANET. In our network and mobility combined design, a user can simulate VANETs in highways with customized road-side and on-board units. Users can create user-defined actions and event handlers to customize simulation scenarios, allowing them to study vehicular traffic, network traffic, or both.

We explain the main components of our design in Section 2 and highlight possible user customizations, such as adding helicopters or embedded highway sensors, in Section 3. In Section 4, we discuss validation of our IDM/MOBIL implementation in ns-3, and in Section 5 we discuss an example of our additions to ns-3. We conclude in Section 6 with a summary and discussion of future work.

## 2 ARCHITECTURE

Here we describe the components of our design, which consists of five main classes (Figure 1):

1. *Vehicle* - a mobile node that contains a wireless communications device
2. *Obstacle* - a *Vehicle* that has no mobility
3. *Model* - the IDM car-following mobility model
4. *LaneChange* – the MOBIL lane change model
5. *Highway* - holds *Vehicle* and *Obstacle* objects and uses a *Vehicle*'s *Model* and *LaneChange* properties to control its mobility

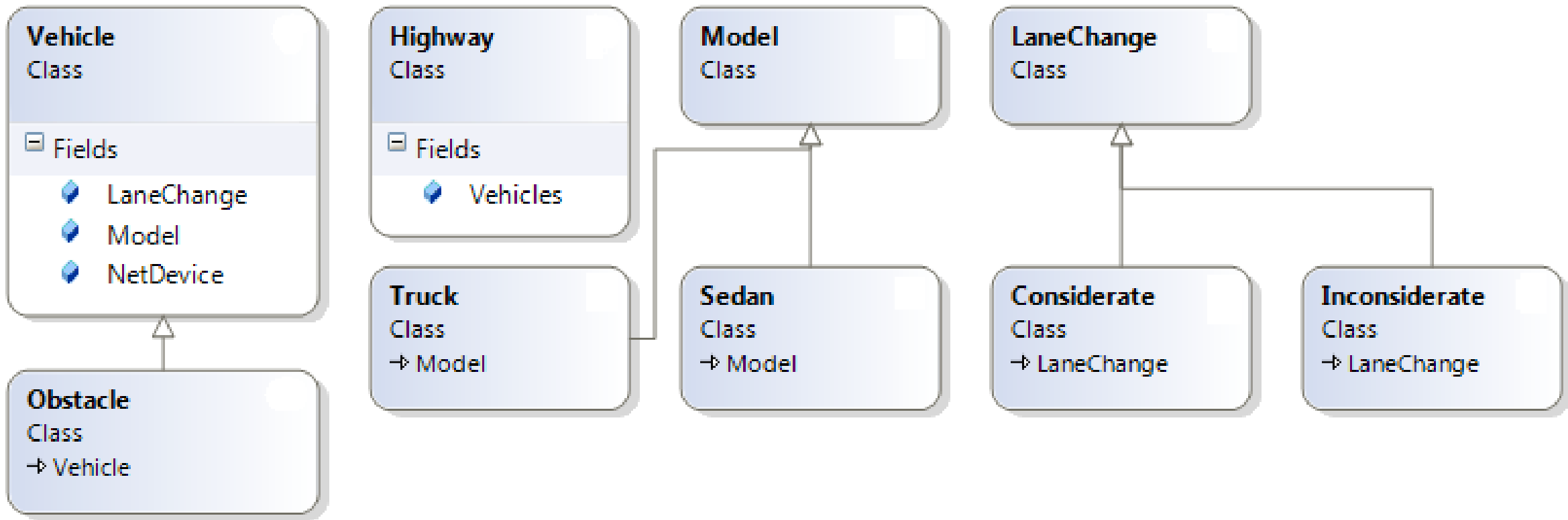

Figure 1: Class diagram of the main components in our design.

*Highway* uses the first four classes to generate the traffic in a highway. Since vehicular mobility models, and especially car-following models like the one we implement, need to know the position and mobility of other vehicles, the *Highway* object must be used to control the mobility of all vehicles. Users can customize *Highway* (including highway length, uni- or bi-directional traffic flow, number of lanes, lane width, and center median width) to create a variety of simulation scenarios.

In the following sections, we will describe each of the classes in order. The source code, including an example and documentation, is available at <cs.odu.edu/vanet/Software/ns3-highway/>.

## 2.1 Vehicle

A *Vehicle* is a mobile node that contains a wireless communications device. A *Vehicle* has the following properties:

- vehicleID
- width – width of the vehicle in meters
- length – length of the vehicle in meters
- lane – lane number on the highway where the vehicle is located
- direction – {-1, 1} (Assume eastbound is 1 and westbound is -1).
- position – a vector ($x$, $y$, $z$), where $x$ is the rear position of the vehicle, $y$ is the center of the vehicle, and $z$ is the altitude of the vehicle above the highway (all units in meters)
- velocity – in m/s
- acceleration – in m/s$^2$
- model – mobility model settings, desired velocity is associated with the mobility model
- lanechange – lane change model settings

In our design, the *Highway* object is in charge of managing the positions, directions, and the lane numbers of its vehicles. A *Vehicle*'s acceleration and velocity can be set manually or can be calculated based on the IDM mobility model rules. A *Vehicle* is able to change lanes, if necessary and if possible, based on the MOBIL lane change model. *Vehicle* objects can either be manually created and inserted onto the *Highway* or can be automatically injected onto the *Highway*.

Since a *Vehicle* contains a wireless communications device, we can control the vehicle's WiFi capabilities. *Vehicles* are able to communicate (send/receive) through the standard ns-3 WiFi channels. The messages, including sent and received packets, and all related events can be captured by setting the appropriate event handlers to the implemented callbacks, which are designed and considered for these purposes. A *Vehicle* can unicast packets or it can send broadcast messages. The user has full control on how to schedule the sending process and how to handle the receive callback. There are also several callbacks for the purpose of tracing the different layers of the network and the mobility of the vehicle. These help the user create and trace simulation scenarios easily.

### 2.2 Obstacle

An *Obstacle* is a static node that contains a wireless communications device. It is inherited from the *Vehicle* class and has all of the capabilities of a *Vehicle* except that it cannot be mobile (i.e., `velocity = acceleration = model = lanechange = 0`). An obstacle can be used as an barrier to close a lane or to temporarily create stoppages that result in congestion on the highway. An obstacle can also be used as a roadside unit or other piece of infrastructure along, but outside of, the highway. If an *Obstacle* is placed on the highway, it must have a direction and lane number. Anything that can be done to a *Vehicle* object can be done to an *Obstacle* object (aside from affecting mobility), so in the rest of this paper we will just use the term *Vehicle*.

### 2.3 Mobility Model

*Model* is the class that implements the mobility model for a *Vehicle*. We have implemented the Intelligent Driver Model (IDM) in ns-3 based on equations and parameters developed by Treiber (Treiber et al. 2000, Treiber 2006a). IDM is a car-following model, meaning that each vehicle's acceleration or deceleration depends upon its own velocity, its desired velocity, and the position and velocity of the vehicle immediately in front in the same lane, which Treiber calls the *front vehicle*.

Each vehicle in IDM has a desired velocity, safety time headway (time needed to cover a gap between two vehicles, e.g., the "2 second rule"), acceleration in free-flow traffic, comfortable braking deceleration, and desired minimum distance to the front vehicle. IDM uses these parameters and the current state of the vehicle and front vehicle to compute the new acceleration. Acceleration is, in turn, used to update the velocity and position of the vehicle. Note that acceleration necessarily decreases towards 0 when the velocity of the vehicle approaches the desired velocity.

The function *CalculateAcceleration* in the *Model* class uses the IDM equations to calculate and return the new acceleration at each time step. The vehicle's new velocity and position are then adjusted based on this new acceleration.

For customizability, each vehicle can have its own set of IDM parameters. Treiber suggests different parameter settings for cars and trucks. For example, trucks have a lower desired velocity, acceleration in free-flow, and comfortable deceleration than cars. "Careful" drivers would have a high safety time headway, and "pushy" drivers would have a low safety time headway, higher desired velocity, acceleration in free-flow, and comfortable deceleration.

In our design, we also allow each *Vehicle* object to have its own IDM parameters. We have included reasonable default values for cars (the *Sedan* class) and trucks (the *Truck* class). The user can create their own vehicle types with different parameter values for specific experiments. For example, a user may want to create a mix of careful and pushy drivers, or include sports cars, police cars, emergency vehicles, and buses, all of which would have very different mobility characteristics.

### 2.4 Lane Change Model

*LaneChange* is the class that implements the lane changing model for a *Vehicle*. We have implemented the MOBIL lane change model based on equations and parameters developed by Treiber (Treiber 2006b, Treiber and Helbing 2002). Each lane change in this model must satisfy both the safety criterion and the incentive criterion. The safety criterion states that the lane change must not cause the vehicle that is being changed in front of (the *back vehicle*) to decelerate unsafely (faster than a certain threshold). The incentive criterion is satisfied if the lane-changing vehicle's advantage is greater than the other vehicles' disadvantages. Note that although the incentive criterion is usually much easier to satisfy than the safety criterion, *both* must hold for the lane change to occur. In addition, the IDM rules still apply, meaning that the new front vehicle must be a certain distance ahead in order for the lane change to occur.

To compute the incentive criterion, MOBIL first calculates the lane-changing vehicle's advantage. This is simply the difference between the vehicle's current acceleration and the vehicle's new acceleration after the lane change. The goal is to increase the acceleration, or to reduce the braking deceleration, which are essentially the same things. The disadvantage to both the back vehicle in the current lane and the back vehicle in the new lane are considered. Again, this is done by comparing the acceleration before the lane change with the acceleration after lane change.

To allow for some variability in how aggressive drivers are in deciding when to change lanes, MOBIL weights the other vehicles' disadvantage with a politeness factor, $p$. When $p \geq 1$, the driver is considerate and puts others' disadvantages equal to or ahead of their own advantage. In reality, most drivers are in the $0 < p \leq 0.5$ range, where some weight is given to other drivers' disadvantage. If $p$=0, the driver is inconsiderate, discounting the disadvantage to others.

MOBIL also includes a right-lane bias parameter when computing the incentive criterion. This parameter allows for modeling situations in countries where passing a vehicle on the right is not allowed. The parameter can also be used to allow vehicles to pass from either side or prevent trucks from travelling in the leftmost lanes.

The function *CheckLaneChange* in our *LaneChange* class returns a boolean to indicate if the lane-change can take place or not. *CheckLaneChange* uses the MOBIL equations and suggested parameters along with the statuses of the lane-changing vehicle, the current front vehicle, the new front vehicle, and the new back vehicle. As with our IDM implementation, we have included reasonable default values for each of these parameters. We provide a *Considerate* driver class and an *Inconsiderate* driver class. The user can, of course, create their own driver types with different parameters.

## 2.5 Highway

*Highway* is the class that holds *Vehicles* and manages their mobility. We will discuss *Highway's* physical properties, *Vehicle* management tasks, and how users can control vehicles on the highway in order to customize simulations.

### 2.5.1 Physical Properties

*Highway* represents a straight highway topology and has the following physical properties:

- length – length of the highway in meters (up to 10,000 m)
- number of lanes – in each direction [1,5]
- lane width – in meters
- median gap – width of the median, in meters
- bidirectional – true if the highway contains two-way traffic, false if the highway is one-way

Figure 2 shows two example highway configurations. Figure 2a is a unidirectional highway with three lanes, and Figure 2b is a bidirectional highway with four lanes in each direction.

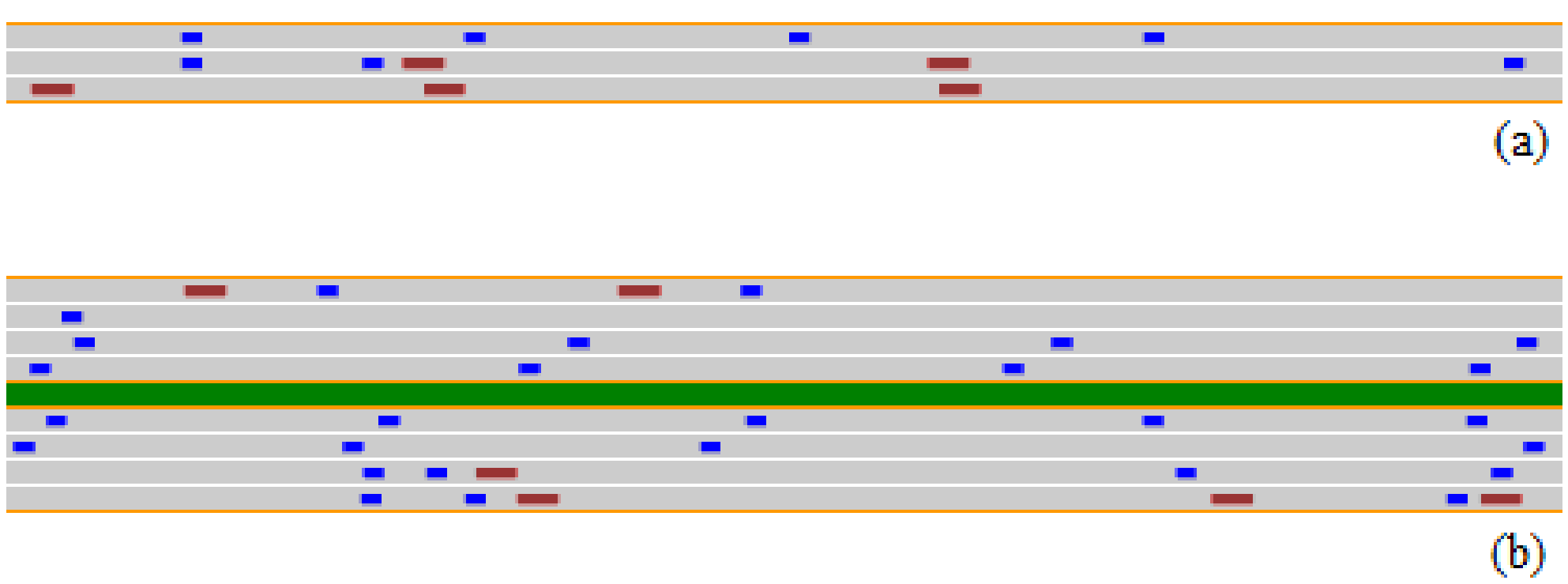

Figure 2: A small segment of a highway. Cars are represented by small blue rectangles, and trucks are represented by larger red rectangles. (a) unidirectional highway with three lanes, (b) bidirectional highway with four lanes in each direction and a separating median.

### 2.5.2 Vehicle Management

There are several *Vehicle* management functions that *Highway* performs. *Highway* can automatically create *Vehicle* objects with certain parameters, automatically insert these created objects into lanes, and move each *Vehicle* according to its mobility and lane change models.

#### 2.5.2.1 Automatic Creation and Injection of Vehicles

When the *AutoInjection* parameter of *Highway* is true, *Vehicle* objects will be automatically created and injected onto the highway. For this purpose, *Highway* creates default mobility models with parameters set appropriately for the standard car and truck, named *SedanModel* and *TruckModel*, respectively. *Highway* also creates default lane change models with appropriate parameters set for cars and trucks. The ratio of cars to trucks that are created is controlled by the *injectionMix* parameter. Automatically-created *Vehicle* objects are provided with default WiFi Phy/Mac settings appropriate for VANETs.

*Highway* stores each lane as a list structure. When a *Vehicle* object is added to *Highway*, it is inserted in its proper place according to its lane, direction, and *x* position. For auto-injection, there is a *minGap* parameter that specifies the minimum distance between two vehicles entering the highway. Newly created *Vehicle* objects are not inserted until the *x* position of the last *Vehicle* in the lane is at least *minGap* meters from the start of the highway. Vehicles are inserted with a negative *x* position, so that the front of the vehicle starts at the start of the highway (`x = 0`). Each lane is checked to see if a *Vehicle* can be added, in round-robin fashion, starting with the rightmost lane (`lane = 0`) in the eastbound direction (`direction = 1`) and ending with the leftmost lane in the westbound direction (`direction = -1`, if using bidirectional traffic). Thus, on a bidirectional highway, vehicles are added to both directions at the same rate.

#### 2.5.2.2 Mobility of Vehicles

Every *DeltaT* seconds, *Highway* calls its *step* function which updates the position, velocity, and acceleration of each *Vehicle* according to its mobility model. In this way, vehicles with different mobility characteristics (e.g., trucks, emergency vehicles) can be represented on the same highway. *Vehicles* are updated by lane in round-robin fashion, starting with the *Vehicles* in the rightmost lane in the eastbound direction. After the update, if a *Vehicle*'s *x* position is greater than the length of the *Highway*, the *Vehicle* is removed from the lane list. After all *Vehicle* positions have been updated, automatic injection of new *Vehicles* occurs.

The opportunity for each vehicle to change lanes is evaluated every 10 * *DeltaT* seconds to prevent unrealistic lane-changing patterns (e.g., vehicles changing lanes multiple times in less than 1 second). If a vehicle can safely change lanes (according to the *Vehicle*'s MOBIL parameters), *Highway* removes the *Vehicle* from the current lane and adds it to the target lane at the *x* position specified according to IDM/MOBIL. When a lane change is allowed, it occurs before mobility is updated, so a *Vehicle* changing lanes only has its mobility updated one time in *DeltaT* seconds.

The best case driver reaction time is 0.7 seconds (Green, 2000). Vehicle positions should be updated more often than the driver reaction time, and we choose 0.1 seconds for the default value of *DeltaT* as a tradeoff between efficiency and accuracy. Reducing *DeltaT* (i.e., having the *step* function called more often) will produce a more detailed translation of the position of the vehicle, but will result in a slower simulation (Figure 3). Increasing *DeltaT* (i.e., having the *step* function called less often) will cause less accuracy in mobility since each step may result in a larger displacement of the vehicles (Figure 4).

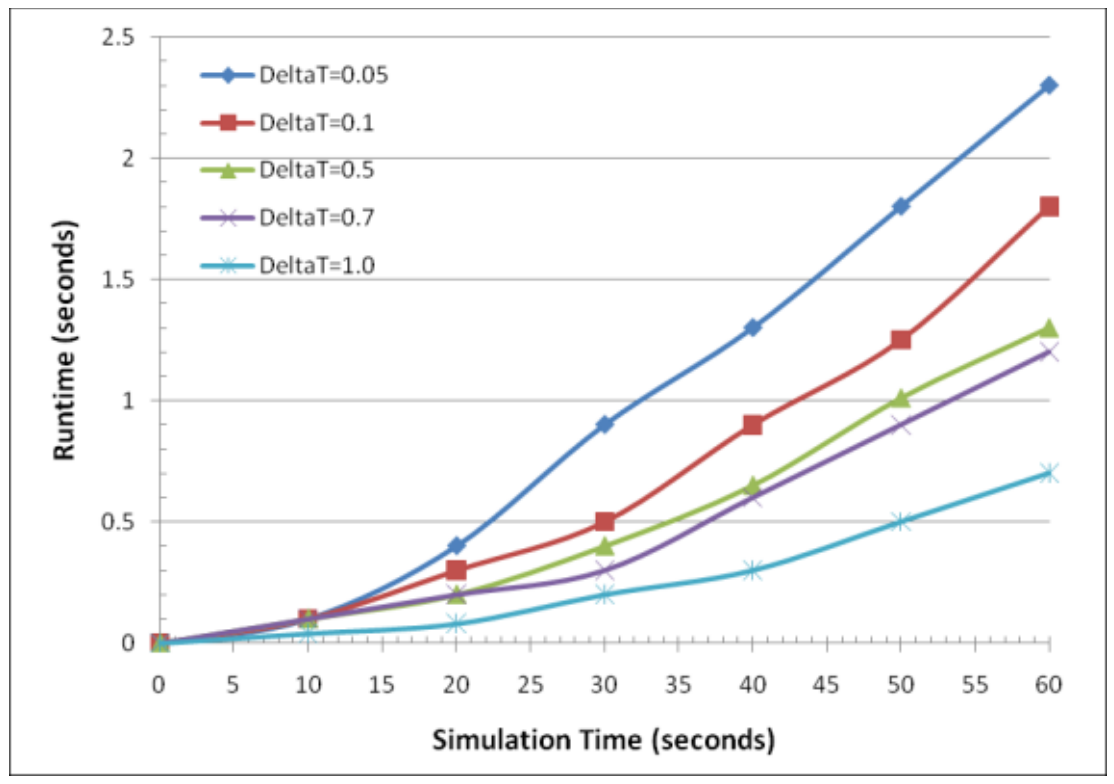

Figure 3: The elapsed real time for 1 minute of dense traffic simulation (average 180 vehicle/km).

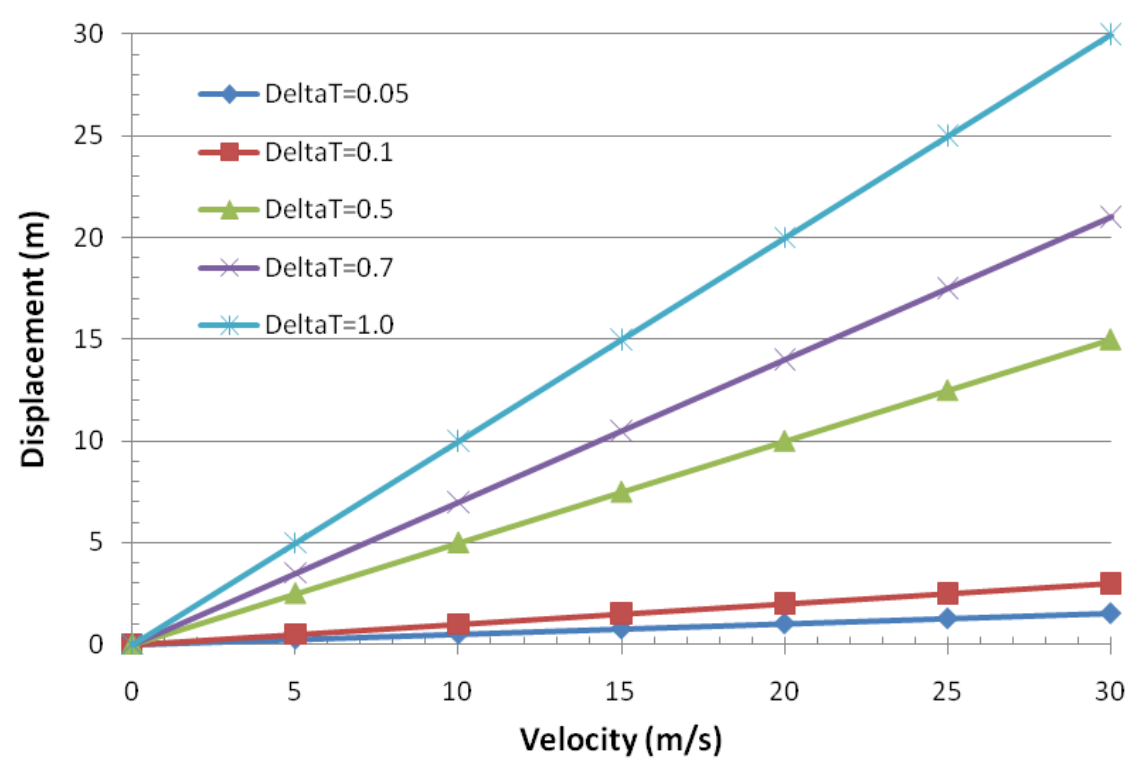

Figure 4: A vehicle's displacement vs. velocity in a single simulation step with different DeltaT values.

### 2.5.3 User Control of Vehicles

To allow for feedback between the network and the mobility model, there must be a way for the user's application code to interact with individual *Vehicle* objects. *Highway* allows the user to access any *Vehicle* object through its *VehicleID* using *FindVehicle()*. The user can then use this object to change any of the *Vehicle*'s parameters. In addition, *Highway* provides *FindVehiclesInRange()* which returns a list of all *Vehicle* objects within *range* meters of the given *Vehicle*. *FindVehiclesInSegment()* returns a list of all *Vehicle* objects in a particular lane between positions $x_1$ and $x_2$. To access these *Vehicle* objects at particular times, *Highway* triggers several events that can be bound to an event handler created by the user. The events *InitVehicle*, *ControlVehicle*, and *ReceiveData* are discussed below. In addition, there are several other events, such as *DevRxTrace* and *PhyRxErrorTrace,* for the purposes of tracing the communication channel, the PHY/MAC layer, and the behavior of the network devices installed on vehicles.

*InitVehicle* is triggered at *Highway* initialization time. This gives the user the ability to create customized scenarios or modify the initial settings. Although the user can create and position *Vehicle* objects at any time, inside this event handler is the ideal place to create and place initial objects on the highway. If *AutoInjection* is set to true in *Highway*, automatically-created *Vehicles* will move around the previously placed *Vehicles*. The event handler is passed a pointer to the *Highway* and a reference to a *vehicleID* (set to 1 initially). Any manually-created *Vehicles* should use and increment this *vehicleID* so that all objects will have unique IDs. Note that any manually-created *Vehicles* will be controlled by *Highway* according to the *Vehicle's* mobility model. The event handler should return *true* if *Vehicles* have been manually added to the *Highway* or default settings have been modified. In this case, *Highway* will sort the lane lists based on the *Vehicle* positions. If no *Vehicles* have been added, there is no reason to sort the lists, so the event handler should return *false*.

For each *Vehicle*, *ControlVehicle* is triggered by the *step* function, which is executed every *DeltaT* seconds. In this way, the user has full control of each *Vehicle* at each time step. For example, a particular *Vehicle* could be made to decelerate or stop in order to create traffic congestion. In addition, this event handler is an ideal place to output the locations of all *Vehicles* in order to produce traffic visualizations. The event handler is passed a pointer to the *Highway*, a pointer to the particular *Vehicle*, and the value of *DeltaT*. If the event handler has changed the *Vehicle*'s position, it should return *true*, so that the *Vehicle*'s acceleration will not be updated by the mobility model. Otherwise, the event handler should return *false* so that *Highway* will adjust the *Vehicle's* position according to its mobility model.

*ReceiveData* is triggered when any *Vehicle* successfully receives data from the network. The event handler is passed a pointer to the *Vehicle* that received the data, a pointer to the data packet, and the address of the packet's sender.

## 3 CUSTOMIZATIONS

We provide a basic framework for a straight highway scenario and tools for generating communicating vehicles traveling with a realistic mobility. There are many possible customizations that can be made using this framework. We describe a few customizations that can be made with *Vehicle* objects.

Any *Vehicle* can be associated with a parameterized mobility or lane-change model. This allows the user to create simulations that contain various types of vehicles. For example, a police car is a vehicle that during a chase has a higher desired speed and acceleration than a normal vehicle. In addition, the user could set the networking parameters such that the police car also has a more powerful transceiver than a normal vehicle. In another instance, a helicopter used to transmit advertisements, warnings, or reports could be simulated as a *Vehicle* with a positive *z* value (altitude). Since the helicopter does not travel on the highway, it should not be added to or managed by *Highway*. Instead, at every time step (i.e., in the *ControlVehicle* event handler), the helicopter's position should be updated manually.

Stationary roadside units, such as digital guides, placed outside the highway can be created using *Obstacles*. As with *Vehicles* that are outside the highway, these devices would should not be added to *Highway*. As another example, a gantry on top of a highway could be represented as an *Obstacle* with a positive *z* value. Sensors under the road could be *Obstacles* with negative *z* values. These devices may have different communications requirements than standard vehicles, so the user is free to adjust the network parameters as well.

## 4 VALIDATION

In this section, we validate our implementation of IDM/MOBIL in ns-3 against Treiber's own implementation of IDM/MOBIL in a Java applet (<http://www.traffic-simulation.de>). The first step is to validate that the functions *Model*::*CalculateAcceleration()* and *LaneChange::CheckLaneChange()* produce output correctly in comparison with Treiber's formula, model, and code individually with various input and mobility model settings. The second step is to produce simple traffic in a one lane roadway and compare the vehicle's acceleration, deceleration, velocity, and position at each simulation interval. Finally, we need to show that despite the difference in our design and the logic of *step* function, we are able to create traffic similar to that created by Treiber's applet.

The first two steps have been performed during code implementation and testing. We omit these for brevity. Here we show the results of the third step of validation. We use Treiber's Java applet to produce traffic on a straight two lane roadway for several traffic inflow rates. We record traffic statistics (simulation time, vehicle type, acceleration, velocity, position, and lane) at two points. Point *A* is the roadway entrance, and point *B* is 500 m from the entrance. We apply the generated traffic recorded at point *A* in Treiber's applet to our ns-3 simulation and record the traffic statistics at point *B*. This is to mitigate the different injection models used by Treiber's applet and our code. We compare the traffic at point *B* in Treiber's applet with the traffic at point *B* in our ns-3 code during a 5 minute simulation. Figure 5 shows the average traffic density over the 500 m as the traffic inflow rate increases and with different desired speeds. The results between the two applications are almost identical. Figure 6 shows the average differences in position and speed between the two applications for each vehicle as it passes point *B*. Again, there is very little difference between the two. The position differences are less than 7 mm, and the speed differences are less than 1 cm/s.

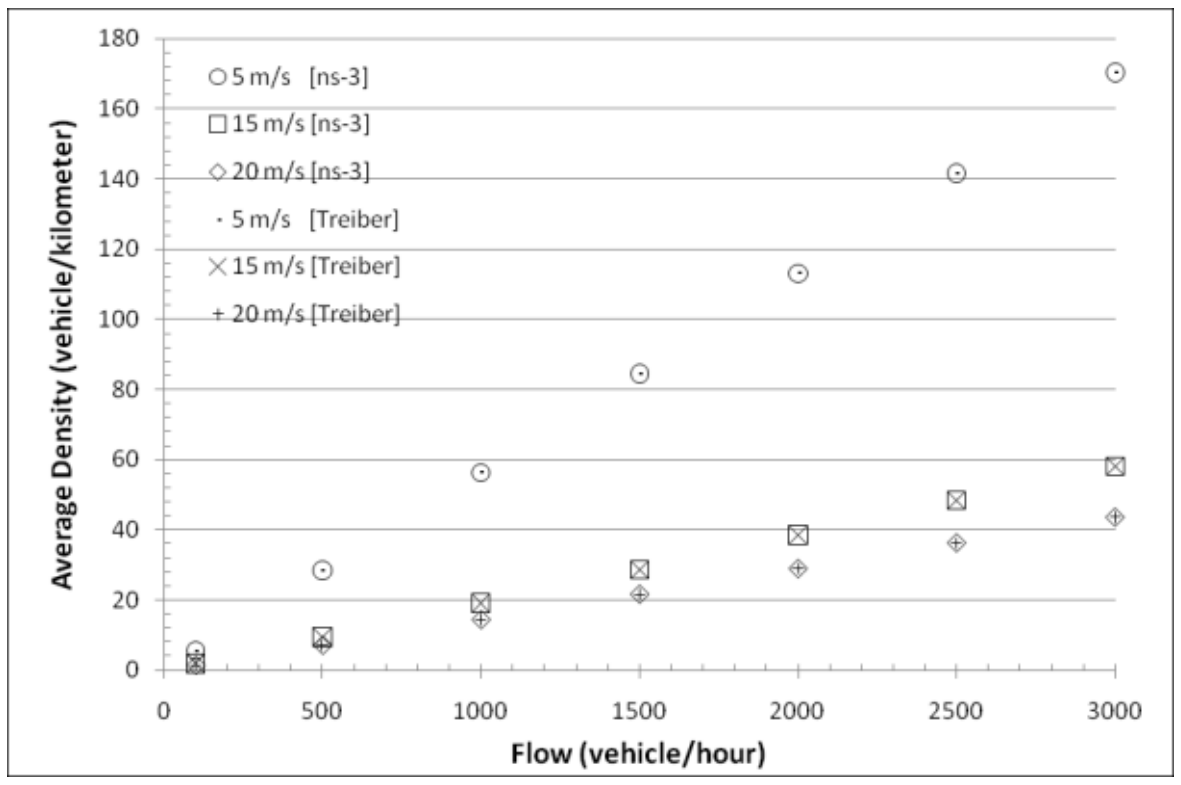

Figure 5: Comparison between average density results of our code in ns-3 and Java applet for different traffic inflow and different desired velocity.

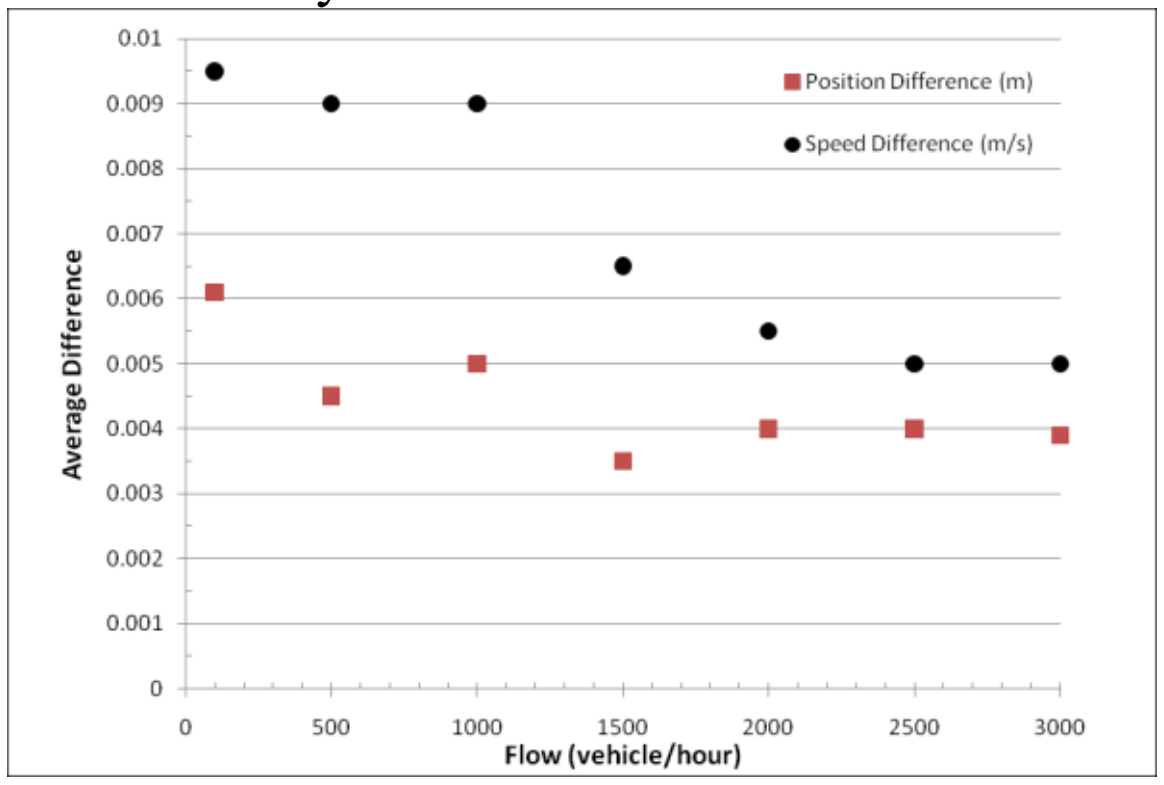

Figure 6: Average difference in position (m) and average speed (m/s) between ns-3 version and Treiber Java applet for different traffic densities.

## 5 EXAMPLE

We have provided an example to show how to create a customized highway, set parameters, handle events, and control which vehicles send and receive customized messages. This example, located at `<http://cs.odu.edu/vanet/Software/ns3-highway/>`, demonstrates how a user can have full control of events to produce the desired scenarios and experiments. The example generates output suitable for plotting vehicle positions using *gnuplot* or other graph-plotting tool.

We have created a *Controller* class to handle events and create special vehicles. The highway is a bidirectional 1 km roadway with two lanes in each direction. The lane width and median width are both 5 meters. The sedan-truck mixture is 80%, so 80% of vehicles are sedans and 20% are trucks. Automatically-generated vehicles will enter and be injected to the highway with at least a 10 meter gap. We place a broken car (*Obstacle* object) in the middle of the highway (`lane=0, direction=1, x=500`) which broadcasts a safety message revealing its location and asking for help every 5 seconds. We also create a police car with a *VehicleID* of 2. The police car is faster than a normal car and has a higher wireless transmission range. It listens for messages and unicasts a reply for each received request. The police car will decelerate when it reaches the broken car and will eventually stop nearby.

The generated output points can be directed to *gnuplot* to be plotted and animated. Figure 7 shows the *gnuplot* snapshot after 2 minutes and 40 seconds of the simulation. The police car reached the broken car at 500 meters after 20 seconds and stopped in the second lane, causing congestion.

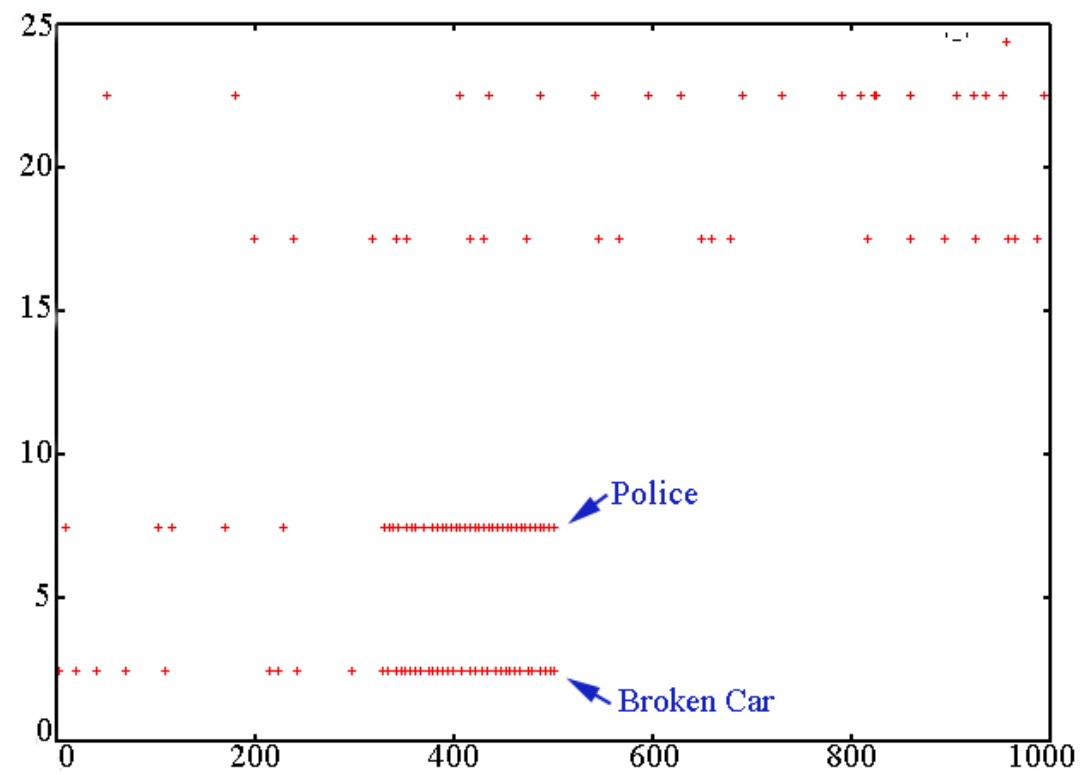

Figure 7: A sample plotted highway output for a 1000m roadway with two lanes in each direction. This snapshot is taken at time 2 minutes, 40 seconds. The police car has stopped in the lane next to the broken car at time 20 seconds, causing the congestion behind it.

Below, we show a skeleton of a *Controller* class and *main*() function. Comments that are shown in italics are placeholders for user-defined code.

**Controller.h**

```
class Controller : public Object
{
    private:
        Ptr<Highway> m_highway;
        // other local variables
    public:
        Controller();
        Controller(Ptr<Highway> highway);
        // event handlers
        bool InitVehicle (Ptr<Highway> highway, int& vehicleID);
        bool ControlVehicle (Ptr<Highway> highway, Ptr<Vehicle> vehicle, double dt);
        void ReceiveData (Ptr<Vehicle> veh, Ptr<const Packet> pckt, Address addr);
        // other function declarations
};
```

**Controller.cc**

```
Controller::Controller() {}
Controller::Controller(Ptr<Highway> highway) {m_highway = highway;}

bool Controller::InitVehicle(Ptr<Highway> highway, int& vehicleID)
{
    // objects to create, settings to change at highway initialization time
    return true; // let Highway sort vehicles in highway lanes
}

bool Controller::ControlVehicle(Ptr<Highway> hw, Ptr<Vehicle> veh, double dt)
{
    // actions that should occur each time this vehicle's mobility is updated
    return false; // let Highway manage the mobility of this vehicle
}

void Controller::ReceiveData(Ptr<Vehicle> veh, Ptr<const Packet> pckt, Address addr)
{
    // actions that should occur each time a message is received by this vehicle
}
```

**main()**

```
int main (int argc, char *argv[])
{
    // Create Highway and Controller
    Ptr<Highway> highway = CreateObject<Highway>();
```

```
    Ptr<Controller> controller = CreateObject<Controller>(highway);

    // Set highway parameters

    // Bind highway events to event handlers in controller
    highway->SetInitVehicleCallback(MakeCallback(&Controller::InitVehicle, controller);
    highway->SetControlVehicleCallback(MakeCallback(&Controller::ControlVehicle, controller);
    highway->SetReceiveDataCallback(MakeCallback(&Controller::ReceiveData, controller);

    // Schedule the highway and run the simulation
    Simulator::Schedule(Seconds(0.0), &Start, highway);
    Simulator::Schedule(Seconds(100.0), &Stop, highway);
    Simulator::Stop(Seconds(100.00));
    Simulator::Run();
    Simulator::Destroy();
    return 0;
}
```

## 6 CONCLUSION AND FUTURE WORK

In this paper, we described the first implementation of a vehicular mobility model integrated with the networking functions in ns-3. Integrated VANET simulators that include both mobility and network models are essential, allowing network communications to affect vehicle mobility, which is one of the main goals of future VANET deployments (e.g., network messages may prompt drivers to slow down early or to take an alternate route). Our implementation allows for this feedback by triggering an event each time a network message is received and each time vehicle mobility is updated. User-created event handlers can then send network messages or alter the mobility of the vehicle in response to the triggered event. These features can facilitate more detailed simulations of VANETs.

Realistic vehicle mobility is achieved through the validated implementation of the IDM car-following model and the MOBIL lane-change model. We introduced the *Highway* class, which not only simulates a straight roadway, but also manages the mobility of all vehicles on the highway. Our implementation also allows the user to take advantage of automatically created and inserted vehicles or to manually insert vehicles at any point along the highway. In addition, our implementation allows for the customization of almost all aspects of the simulation so that the research can study a wide variety of scenarios.

In future work, we plan to extend our implementation for urban areas (intersections) and add the ability to read in and use detailed maps instead of a single straight highway. We also have plans to implement and develop the WAVE/DSRC standard in ns-3. This will allow users to simulate realistic wireless communication for VANETs based on the standard, which includes multi-channel operation. We hope that this addition to ns-3 along with our future work will allow researchers to easily perform high-quality VANET simulations.

## AUTHOR BIOGRAPHIES

HADI ARBABI is a PhD candidate in Computer Science at Old Dominion University. He received his B.S. in Computer Engineering from Shiraz University in 2001. He received his M.S. in Computer Science from ODU under the supervision of Dr. Stephan Olariu in 2007 and is pursuing his PhD at ODU under the supervision of Dr. Michele C. Weigle. His PhD focus is on VANETs, specifically their application to monitoring roadway traffic. His email address is <marbabi@cs.odu.edu>.

MICHELE C. WEIGLE is an Assistant Professor of Computer Science at Old Dominion University. She received her Ph.D. from the University of North Carolina at Chapel Hill in 2003. Her research interests include vehicular networks, wireless and mobile networks, network protocol evaluation, network simulation and modeling, and Internet congestion control. She is a member of ACM, ACM SIGCOMM, ACM SIGMOBILE, IEEE, and IEEE ComSoc. Her e-mail address is <mweigle@cs.odu.edu>.